\documentclass[conference]{IEEEtran}
\IEEEoverridecommandlockouts
\usepackage{cite}
\usepackage{amsmath,amssymb,amsfonts}
\usepackage{algorithm}
\usepackage{algpseudocode}
\usepackage{graphicx}
\usepackage{textcomp}
\usepackage{xcolor}
\newcounter{MYtempeqncnt}

\def\BibTeX{{\rm B\kern-.05em{\sc i\kern-.025em b}\kern-.08em
    T\kern-.1667em\lower.7ex\hbox{E}\kern-.125emX}}
\begin{document}
\bstctlcite{BSTcontrol}
\title{Fast Beam Training and Performance Analysis for Extremely Large Aperture Array \thanks{This work was supported by the National Key R\&D Project of China under Grant No. 2023YFB2905000. (\textit{Corresponding Author: Hongkang Yu})}}

\author{%
  Yuan Si\IEEEauthorrefmark{1}\IEEEauthorrefmark{2},
  Hongkang Yu\IEEEauthorrefmark{1}\IEEEauthorrefmark{2},
  Yijian Chen\IEEEauthorrefmark{1}\IEEEauthorrefmark{2},
  \\
  \IEEEauthorblockA{%
    \IEEEauthorrefmark{1}State Key Laboratory of Mobile Network and Mobile Multimedia Technology, Shenzhen 518055, China
  }
  \IEEEauthorblockA{%
    \IEEEauthorrefmark{2}Wireless and Computing Product R\&D Institute, ZTE Corporation, Xian 710114,  China
  }
  \{si.yuan, yu.hongkang, chen.yijian@zte.com.cn\}
}
\maketitle

\begin{abstract}
	Extremely large aperture array (ELAA) can significantly enhance beamforming gain and spectral efficiency. Unfortunately, the use of narrower beams for data transmission results in a substantial increase in the cost of beam training. In this paper, we study a high-efficiency and low-overhead scheme named hash beam training. Specifically, two improved hash codebook design methods, random and fixed, are proposed. Moreover, we analyze beam alignment performance. Since the derived beam alignment success probability is a complex function, we also propose a heuristic metric to evaluate the impact of codebook parameter on performance. Finally, simulation results validate the theoretical analysis, indicating that the proposed beam training scheme can achieve fast beam alignment with lower overhead and higher accuracy.
\end{abstract}

\section{Introduction}
\IEEEPARstart{W}{ith} the development of wireless communication, the operating frequency becomes higher, and the array aperture becomes larger. 5G new radio (NR), which uses massive multi-input multi-output (MIMO) as one of the key technologies, has applied the millimeter-wave band and supports 32 antenna ports with hundreds of antennas \cite{ref1}. In the future, 6G will further introduce terahertz (THz) wave band and the extremely large aperture array (ELAA) equipped with thousands of antennas. More antennas and larger array aperture can bring higher beamforming gain to combat the severe path loss in high frequency bands and ensure the coverage of the base station \cite{ref3}.

However, higher beam gain leads to narrower beams, which makes the overhead of beam training overwhelming. How to achieve fast beam alignment with lower overhead is a major challenge. The traditional beam sweeping scheme sends training beams sequentially, which brings high overhead, is not suitable for the ELAA scenario. To solve this, hierarchical beam training schemes were proposed \cite{ref4,ref5,ref7}, where the base station first sends few wide beams, narrows the beam search range according to the feedback from the user, and finally achieves beam alignment with specific resolution. Unfortunately, this scheme can reduce overhead only in the case of few users, and requires multiple feedbacks, which increases the delay.

Channel estimation based on compressive sensing is another scheme that can achieve high-precision beam alignment \cite{ref8, ref9}. Taking advantage of the sparsity  of the channel in the angular domain at high frequency bands, we can obtain the estimated channel vector by the sparse signal recovery algorithm with a small amount of pilot overhead. However, the low overhead comes at the expense of robustness and increased computational complexity. Moreover, the channel estimation performance deteriorates rapidly as the signal-to-noise ratio (SNR) decreases.

Recently, a high-efficiency scheme named hash beam training received extensive attention, where the training beams comprise multiple main lobes with different directions \cite{ref11, ref12}. Based on the measurement results, the user can infer whether the optimal beam direction is among these specific main lobes. Through cross-validation of multiple measurement results, the best beam direction can be selected and fed back to the base station. Hash beam training not only saves training overhead but also has lower computational complexity, enabling better support for multi-user scenarios and applicable in both ELAA and reconfigurable intelligent surfaces (RIS). The design of the training beam codebook has a significant impact on the performance of the hash beam training scheme. Unfortunately, existing hash codebooks are generated randomly without specialized design, so there is still a large space for improvement.

Addressing the aforementioned issues, this paper focuses on the codebook design in hash beam training, proposing two improved codebook design schemes, random and fixed, and conducts a theoretical analysis of the beam alignment success probability. Due to the complex form of the derived theoretical results, a novel heuristic metric with closed-form expressions is also proposed to evaluate the impact of hash codebooks on beam alignment performance. According to this parameter, the key parameters used to generate the hash codebook can be quickly determined. Finally, simulation results verify the correctness of the theoretical analysis and show better performance than the existing schemes.


\section{System model}\label{sec2}
Consider a THz point-to-point multi-input single-output (MISO) system, where the base station is equipped with a uniform linear array (ULA) of $N$ antennas to serve a single-antenna user. Assuming that the user position is constant or keep moving at a low speed, the channel has the block fading property, that is, it can be regarded as constant over a period of time. Considering the sparsity of the THz channel, we use the geometric channel model \cite{ref3}, which can be expressed by 
\begin{equation}
\bold{h}=\sum^P_{p=1} \alpha_p \bold{a}(\theta_p),
\end{equation}
where $\bold{a}(\theta)=\frac{1}{\sqrt{N}}[1,e^{-j\pi\sin\theta},\dots,e^{-j\pi(N-1)\sin\theta}]^{T}\in\mathbb{C}^{N\times1}$  represents the array response vector, $P$ denotes the number of paths, $\alpha_p$ and $\theta_p$ are the gain and angle of the $p$-th path, respectively. For THz band, the value of $P$ is usually small, e.g., 2 to 3 \cite{ref3}.

We assume that the purpose of beam training is to select an optimal beam from $N$ discrete Fourier transform (DFT) beams for data transmission, the beamforming weight of the $n$-th $(n=1,2,\dots,N)$ DFT beam can be expressed as
\begin{equation}
\bold{f}_n=\bold{a}(\theta_n),
\end{equation}
where $\mathrm{sin}\theta_n=-1+2n/N$.

\begin{figure}[t]
	\centering
	\includegraphics[width=2.8in]{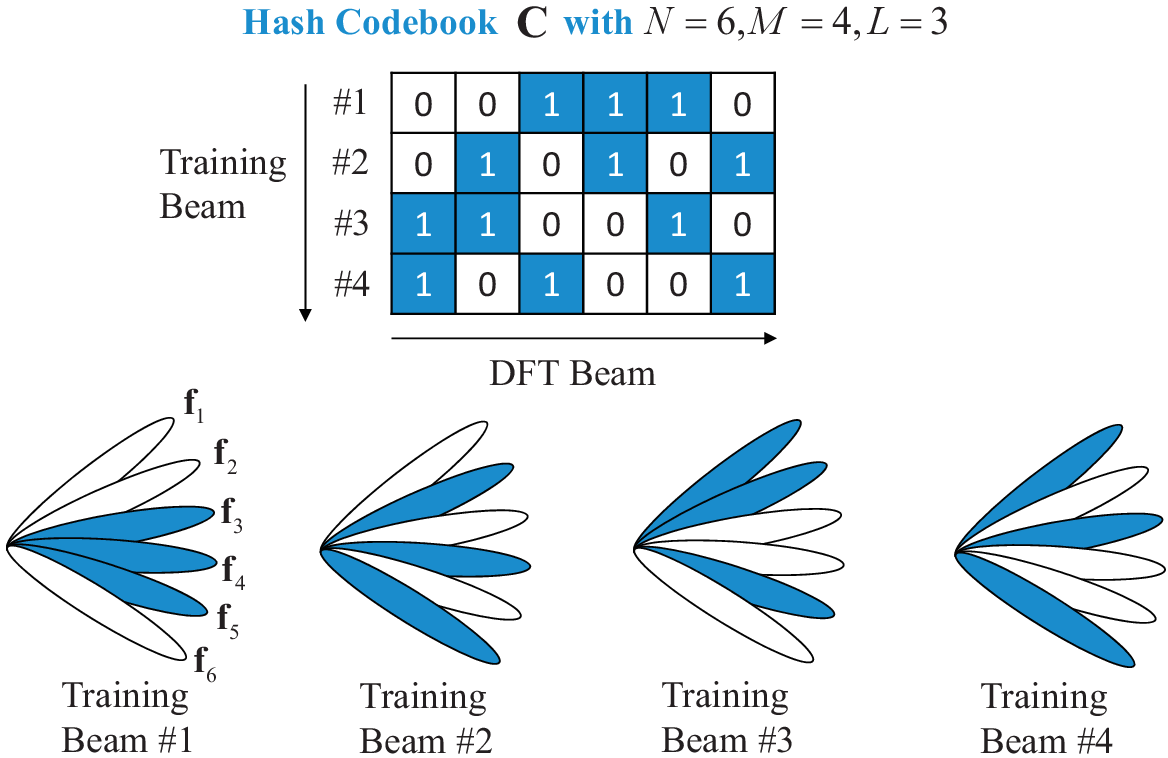}
	\caption{An illustration of hash beam training scheme.}
	\label{fig1}
\end{figure}

In hash beam training scheme, the base station sequentially sends $M$ training beams, each of which is a superposition of $L$ DFT beams, as shown in Fig. \ref{fig1}. The indices of the selected $L$ DFT beams are randomly generated by a hash function, and we use the hash codebook $\mathbf{C}\in\mathbb{C}^{M\times N}$ to describe this process. If $c_{m,n}=1$, it means that the $n$-th DFT beam $\bold{f}_n$ is selected by the $m$-th training beam, while $c_{m,n}=0$ denotes that the $n$-th DFT beam is not selected. Each training beam selects $L$ DFT beams, i.e., $\sum^N_{n=1}c_{m,n}=L$, and the beamforming weight of the $m$-th training beam can be expressed as 
\begin{equation}
\bold{w}_m=\frac{1}{\sqrt{L}}\sum^N_{n=1}c_{m,n}\bold{f}_n.
\end{equation}
Based on the above model, the reference signal received power (RSRP) of the user in the $m$-th training can be expressed as

\begin{equation}\label{eq4}
y_m=\left| \bold{h}^H\bold{w}_m \right|^2+n_m,
\end{equation}
where $n_m\sim \mathcal{N}(0,\sigma^2)$ denotes Gaussian noise.

The advantage of hash-based beam training is that the user can identify the best DFT beam based only on the RSRP $\{y_1,y_2,\dots,y_M\}$, which leads to the low processing complexity. Since the same hash function is preset in the user and the base station, the user can obtain the hash codebook $\bold{C}$ according to the training beam index $m$. After that, users select the best beam by voting mechanism. If the $m$-th training contains the $n$-th DFT beam, the received power $y_m$ of the $m$-th training is accumulated to the measured result $p_n$ of the DFT beam $n$, then the measurement result of the $n$-th beam $p_n$ can be expressed as
\begin{equation}\label{eq5}
p_n=\sum^M_{m=1}c_{m,n}y_m.
\end{equation}
Finally, the user selects the DFT beam $\bold{f}_{\hat{n}}$ with the largest measurement result for reporting, where $\hat{n}=\arg\mathop{\max}\limits_{n} p_n$. 

\textbf{Comment 1}: Many existing beam training schemes can be regarded as special cases of hash beam training, and the difference is only in the hash codebook. For example, the beam sweeping scheme can be represented by a hash codebook $\bold{C}=\bold{I}$ with $M=N,L=1$. The hierarchical beam training scheme in \cite{ref13} corresponds to a hash codebook with $M=2\log_{2}N,L=N/2$. Therefore, the hash codebook design has a significant impact on the beam alignment performance as well as the training overhead, which will be analyzed in the following two sections.

\section{Hash Codebook Design}\label{sec3}
This section proposes several improved codebook design methods, incorporating both random and fixed approaches. In the existing scheme described in \cite{ref11}, each training beam independently selects DFT beams. Given the limited number of training beams, there is a significant variance in the selection frequency of individual DFT beams ${G_n} = \sum\nolimits_{m = 1}^M {{c_{m,n}}} $. As a result, when users are located in different positions, the probability of successful selection for each DFT beam varies considerably, leading to a decrease in overall beam alignment performance.
\subsection{The Proposed Random Hash Codebook}
\label{proposed random}
Addressing the issues present in the existing scheme, the proposed random hash codebook selects DFT beams through a collaborative approach for each training beam, with the goal of equalizing the selection frequency $G_n$ of each DFT beam. To achieve this objective, we constrain the codebook parameters such that $N$ is divisible by $L$, and $M$ is divisible by $N/L$. We divide the training beams into $ML/N$ groups, each containing $N/L$ training beams. Within the same group, training beams randomly select DFT beams through a sampling without replacement, ensuring that each DFT beam appears only once in a training beam group, thereby satisfying ${G_n} = ML/N,\forall n$. The detailed process of the proposed scheme is summarized in Algorithm \ref{alg1}.
\begin{algorithm}[h]
	\caption{Proposed Random Hash Codebook}
	\label{alg1}
	\begin{algorithmic}[1]
		\renewcommand{\algorithmicrequire}{\textbf{Input:}}
		\renewcommand{\algorithmicensure}{\textbf{Output:}}
		\Require {Number of DFT beams $N$, number of training beams $M$, codebook parameter $L$} \Ensure {Hash codebook $\mathbf C$}
		\State Initialize ${\mathbf{C}} = {{\mathbf{0}}_{M \times N}}$
		\For{$m=1,2,\dots,M$}
		\If {$\bmod \left( {m - 1,N/L} \right) =  = 0$}
		\State $\mathcal{S} = \left\{ {1,2, \dots ,N} \right\}$
		\EndIf
		\State Randomly select $L$ elements from $\mathcal S$ to form set $\mathcal{T}$
		\State Set ${c_{m,n}} = 1$ for $n\in\mathcal T$
		\State $\mathcal{S} = \mathcal{S} - \mathcal{T}$	
		\EndFor
	\end{algorithmic}
\end{algorithm}

\subsection{The Proposed Fixed Hash Codebook}
When considering system implementation, the use of random hash codebooks leads to increased complexity. For example, base stations and users need to pre-configure the same random seed for codebook generation. To address this, this subsection introduces the design of a fixed hash codebook. 

Since the codebook is predetermined in actual scenarios, it is feasible to search for the optimal codebook for beam alignment performance using offline simulation. Specifically, we determine the user distribution based on actual scenarios to obtain channel samples, and utilize the method described in Section \ref{proposed random} to generate $X$ proposed random hash codebooks $\left\{ {{{\mathbf{C}}_1},{{\mathbf{C}}_2}, \ldots ,{{\mathbf{C}}_X}} \right\}$. Subsequently, we perform Monte Carlo simulations on these $X$ hash codebooks under identical channel samples, and select the one with the highest beam alignment accuracy as the fixed hash codebook.

Compared to randomly generated hash codebooks, the proposed fixed hash codebook is not only easier to implement in the system but also better matches the actual environment. Related simulation results will be presented in Section \ref{sim}.

\begin{figure*}[t]
	\normalsize
	\setcounter{MYtempeqncnt}{\value{equation}}	
	\setcounter{equation}{13}
	\begin{equation}\label{top1}
		\begin{split}
			\mathbb{P}\left\{ {{G_{{n^{*}}}} = {g_{{n^{*}}}},{G_{{n^{'}}}} = {g_{{n^{'}}}},{K_{{n^{*}},{n^{'}}}} = {k_{{n^{*}},{n^{'}}}},} \right\} &= \\ \mathbb{P}\left\{ {{G_{{n^{*}}}} = {g_{{n^{*}}}}} \right\}\mathbb{P}\left\{ {{K_{{n^{*}},{n^{'}}}} = {k_{{n^{*}},{n^{'}}}}|{G_{{n^{*}}}} = {g_{{n^{*}}}}} \right\}\mathbb{P}&\left\{ {{G_{{n^{'}}}} = {g_{{n^{'}}}}|{G_{{n^{*}}}} = {g_{{n^{*}}}},{K_{{n^{*}},{n^{'}}}} = {k_{{n^{*}},{n^{'}}}}} \right\}
		\end{split}
	\end{equation}
	
	\begin{equation}\label{final_result1}
		\begin{split}
			\tilde p = \sum\limits_{{g_{{n^{*}}}} = 0}^M {\sum\limits_{{k_{{n^{*}},{n^{'}}}} = 0}^{{g_{{n^{*}}}}} {\sum\limits_{{g_{{n^{'}}}} = {k_{{n^{*}},{n^{'}}}}}^M {\mathbb{P}\left\{ {{p_{{n^{*}}}} > {p_{{n^{'}}}}|{\mathbf{C}}} \right\}\mathbb{P}\left\{ {{G_{{n^{*}}}} = {g_{{n^{*}}}},{G_{{n^{'}}}} = {g_{{n^{'}}}},{K_{{n^{*}},{n^{'}}}} = {k_{{n^{*}},{n^{'}}}},} \right\}} } } 
		\end{split}
	\end{equation}
	
		\begin{equation}\label{top2}
		\begin{split}
			\mathbb{P}\left\{ {{K_{{n^{*}},{n^{'}}}} = {k_{{n^{*}},{n^{'}}}}} \right\} =  C_G^{{k_{{n^{*}},{n^{'}}}}}\left( {1 - \frac{{ML - NG}}{N}} \right)\left( {1 - \frac{{ML - NG}}{{N - 1}}} \right){\left( {\frac{{L - 1}}{{N - 1}}} \right)^{{k_{{n^{*}},{n^{'}}}}}}{\left( {1 - \frac{{L - 1}}{{N - 1}}} \right)^{G - {k_{{n^{*}},{n^{'}}}}}}, \\
			{k_{{n^{*}},{n^{'}}}} = 0,1, \ldots ,G.
		\end{split}
	\end{equation}

	\setcounter{equation}{\value{MYtempeqncnt}}
	\hrulefill
	\vspace*{4pt}	
\end{figure*}

\section{Beam Alignment Performance Analysis}\label{sec4}
This section analyzes the impact of hash codebooks on the beam alignment performance from a theoretical perspective. For the sake of simplicity, it is assumed that the channel contains only line-of-sight (LoS) components, i.e., $P=1$. This assumption is reasonable for high-frequency channels, as the gain of non-LoS (NLoS) paths is much smaller than that of LoS paths, having a minor impact on the received power. Furthermore, without loss of generality, we assume that the path gain satisfies $\alpha=1$.
\subsection {Accurate Result of Beam Alignment Success Probability}
Denoting the index of the optimal DFT beam as ${n^{*}}$, the probability of successful beam alignment can be represented as
\begin{equation}
\mathbb{P}\left\{ {{n^{*}} = \hat n} \right\} = \mathbb{P}\left\{ {\mathop  \cap \limits_{{n^{'}} \ne {n^{*}}}  {{p_{n^*}} > {p_{{n^{'}}}}}} \right\},
\end{equation}
where we focus on analyzing the probability of ${p_{{n^{*}}}} > {p_{{n^{'}}}}$. For the LoS channel ${\mathbf{h}} = {\mathbf{a}}\left( \theta  \right)$, ${\left| {{{\mathbf{h}}^{\text{H}}}{{\mathbf{f}}_n}} \right|^2}$ assumes a larger value only when $\theta$ is within the coverage range of the DFT beam. To simplify the derivation, it is further assumed that $\theta$ is located on a discrete grid point, i.e., $\theta  =  - 1 + 2n/N,n = 1,2, \dots ,N$, in which case ${n^{*}} = n$, and ${\left| {{{\mathbf{h}}^{\text{H}}}{{\mathbf{w}}_m}} \right|^2} = {c_{m,{n^{*}}}}/L$. Combining \eqref{eq4} and \eqref{eq5}, we have
\begin{eqnarray}\label{eq7}
	\begin{aligned}
		{p_{{n^*}}} =& \frac{{{G_{{n^{*}}}}}}{L} + \sum\limits_{m = 1}^M {{c_{m,{n^{*}}}}{n_m}}, \\
		 {p_{{n^{'}}}} =& \frac{{{K_{{n^{*}},{n^{'}}}}}}{L} + \sum\limits_{m = 1}^M {{c_{m,{n^{'}}}}{n_m}} ,\forall {n^{'}} \ne {n^{*}},
	\end{aligned}
\end{eqnarray}
where ${K_{{n^{*}},{n^{'}}}} = \sum\nolimits_{m = 1}^M {{c_{m,{n^{*}}}}{c_{m,{n^{'}}}}}$ represents the number of times DFT beams ${n^{*}}$ and ${n^{'}}$ are simultaneously selected by the training beams. According to \eqref{eq7}, 
\begin{equation}\label{eq9}
	{p_{{n^{*}}}} - {p_{{n^{'}}}} = \frac{{{G_{{n^{*}}}} - {K_{{n^{*}},{n^{'}}}}}}{L} + \sum\limits_{m = 1}^M {\left( {{c_{m,{n^{*}}}} - {c_{m,{n^{'}}}}} \right){n_m}},
\end{equation}
we can deduce that $\mathop  \cap \limits_{{n^{'}} \ne {n^{*}}}  {{p_{{n^{*}}}} > {p_{{n^{'}}}}} $ implies
\begin{equation}
	{{\mathbf{\tilde A}}_{{n^{*}}}}{\mathbf{n}} \prec {{\mathbf{\tilde t}}_{{n^{*}}}},
\end{equation}
where ${{\mathbf{\tilde A}}_{{n^{*}}}}$ and ${{\mathbf{\tilde t}}_{{n^{*}}}}$ are respectively the results of removing the ${n^{*}}$-th column of ${{\mathbf{A}}_{{n^{*}}}} \in {\mathbb{R}^{N \times M}}$ and the ${n^{*}}$-th row of ${{\mathbf{t}}_{{n^{*}}}} \in {\mathbb{R}^{N \times 1}}$, with the $(n,m)$-th element of ${{\mathbf{A}}_{{n^{*}}}}$ being ${c_{m,n}} - {c_{m,{n^{*}}}}$, ${{\mathbf{t}}_{{n^{*}}}} = {\left[ {\frac{{{G_{{n^{*}}}} - {K_{{n^{*}},1}}}}{L},\frac{{{G_{{n^{*}}}} - {K_{{n^{*}},2}}}}{L}, \ldots ,\frac{{{G_{{n^{*}}}} - {K_{{n^{*}},N}}}}{L}} \right]^{\text{T}}}$, and ${\mathbf{n}} = {\left[ {{n_1},{n_2}, \ldots ,{n_M}} \right]^{\text{T}}}$.

It should be noted that although both ${{\mathbf{\tilde A}}_{{n^{*}}}}$ and ${{\mathbf{\tilde t}}_{{n^{*}}}}$ are known variables and the elements in $\bf n$ follow mutually independent Gaussian distribution, the distribution of the random variable ${{\mathbf{\tilde A}}_{{n^{*}}}}{\mathbf{n}}$ does not exist mathematically. This is because in hash beam training, it is usually the case that $N>M$, resulting in the covariance matrix ${\mathbf{\Sigma }} = {{\mathbf{\tilde A}}_{{n^{*}}}}{\mathbf{\tilde A}}_{{n^{*}}}^{\text{H}}$ not being full-rank \cite{ref14}. By adding a regularization term, it can be approximated that ${{\mathbf{\tilde A}}_{{n^{*}}}}{\mathbf{n}} \sim \mathcal{N}\left( {{\mathbf{0}},{\mathbf{\Sigma }} + \lambda {\mathbf{I}}} \right)$, and the probability of successful beam alignment $\mathbb{P}\left\{ {{n^{*}} = \hat n} \right\}$ corresponds to the cumulative distribution function (CDF) value at ${{\mathbf{\tilde t}}_{{n^{*}}}}$. Unfortunately, the calculation of CDF value can only be done numerically, which prevents further analysis of the impact of codebook parameters on beam alignment performance.

\subsection{Heuristic Beam Alignment Performance Metrics}
To determine the optimal codebook parameter, i.e., the number of DFT beams $L$ selected by each training beam, this subsection proposes a heuristic metric for evaluating beam alignment performance. Since a fixed codebook can be optimized through extensive offline simulations, the focus here is on identifying the optimal parameter  for random codebooks. The novel metric $\tilde p$ we proposed in this study is the probability that the measurement result ${p_{{n^{*}}}}$ of the optimal DFT beam exceeds that of any other randomly selected beam measurement result ${p_{{n^{'}} \ne {n^{*}}}}$ when generating a random hash codebook.

According to \eqref{eq9}, when the hash codebook $\mathbf C$ is given, $${p_{{n^{*}}}} - {p_{{n^{'}}}}\sim\mathcal{N}\left( {\frac{{{G_{{n^{*}}}} - {K_{{n^{*}},{n^{'}}}}}}{L},\left( {{G_{{n^{*}}}} + {G_{{n^{'}}}} - 2{K_{{n^{*}},{n^{'}}}}} \right){\sigma ^2}} \right),$$ we have
\begin{eqnarray}\label{approx problem}
	\begin{aligned}
	& \mathbb{P}\left\{ {{p_{{n^{*}}}} > {p_{{n^{'}}}}|{\mathbf{C}}} \right\} \\
	& = \frac{1}{2}\left\{ {1 - {\text{erf}}\left( {\frac{{{G_{{n^{*}}}} - {K_{{n^{*}},{n^{'}}}}}}{{L\sqrt {2\left( {{G_{{n^{*}}}} + {G_{{n^{'}}}} - 2{K_{{n^{*}},{n^{'}}}}} \right){\sigma ^2}} }}} \right)} \right\},
	\end{aligned}
\end{eqnarray}
where the values of ${G_{{n^{*}}}}$ and ${K_{{n^{*}},{n^{'}}}}$ are determined by the codebook parameter $L$. Next, we will discuss both existing  and the proposed random hash codebooks.

\subsubsection{Existing random hash codebook}
Utilizing combinatorial mathematics and classical probability theory, the probability of DFT beam ${n^{*}}$ being selected ${g_{{n^{*}}}}$ times is
			
\begin{eqnarray}\label{eq11}
	\begin{aligned}
		&\mathbb{P}\left\{ {{G_{{n^{*}}}} = {g_{{n^{*}}}}} \right\} = C_M^{{g_{{n^{*}}}}}{\left( {\frac{L}{N}} \right)^{{g_{{n^{*}}}}}}{\left( {1 - \frac{L}{N}} \right)^{M - {g_{{n^{*}}}}}},\\
	&	{g_{{n^{*}}}} = 0,1, \dots ,M.
	\end{aligned}
\end{eqnarray}
Given ${G_{{n^{*}}}} = {g_{{n^{*}}}}$, the probability of DFT beams ${n^{*}}$ and ${n^{'}}$ both occurring ${g_{{n^{*}}}}$ times is
\begin{eqnarray}
	\begin{aligned}
		&\mathbb{P}\left\{ {{K_{{n^{*}},{n^{'}}}} = {k_{{n^{*}},{n^{'}}}}|{G_{{n^{*}}}} = {g_{{n^{*}}}}} \right\} =\\
		 &C_{{g_{{n^{*}}}}}^{{k_{{n^{*}},{n^{'}}}}}{\left( {\frac{{L - 1}}{{N - 1}}} \right)^{{k_{{n^{*}},{n^{'}}}}}}{\left( {1 - \frac{{L - 1}}{{N - 1}}} \right)^{{g_{{n^{*}}}} - {k_{{n^{*}},{n^{'}}}}}},\\
		&	{k_{{n^{*}},{n^{'}}}} = 0,1, \ldots ,{g_{{n^{*}}}}. 
	\end{aligned}
\end{eqnarray}
Under the aforementioned conditions, the probability of DFT beam ${n^{'}}$ being selected ${g_{{n^{'}}}}$ times is
\begin{eqnarray}\label{eq13}
	\begin{aligned}
		& \mathbb{P}\left\{ {{G_{{n^{'}}}} = {g_{{n^{'}}}}|{G_{{n^{*}}}} = {g_{{n^{*}}}},{K_{{n^{*}},{n^{'}}}} = {k_{{n^{*}},{n^{'}}}}} \right\} = \\
		&C_{M - {k_{{n^{*}},{n^{'}}}}}^{{g_{{n^{'}}}} - {k_{{n^{*}},{n^{'}}}}}{\left( {\frac{L}{{N - 1}}} \right)^{{g_{{n^{'}}}} - {k_{{n^{*}},{n^{'}}}}}}{\left( {1 - \frac{L}{{N - 1}}} \right)^{M - {g_{{n^{'}}}}}},\\
		&{g_{{n^{'}}}} = {k_{{n^{*}},{n^{'}}}},{k_{{n^{*}},{n^{'}}}} + 1, \ldots ,M. 
	\end{aligned}
\end{eqnarray}
Combining \eqref{eq11}-\eqref{eq13}, we can obtain the proposed heuristic metric $\tilde{p}$ for the existing random hash codebook. The result \eqref{top1} and \eqref{final_result1} is shown at the top of this page. 
\subsubsection{Proposed random hash codebook}
\setcounter{equation}{16}
For the proposed random codebook, ${G_n} = G = ML/N,\forall n$, it is only necessary to analyze the distribution of  ${K_{{n^{*}},{n^{'}}}}$, which is shown in \eqref{top2} at the top of this page. Finally,  the proposed heuristic metric $\tilde{p}$ for the proposed random hash codebook can be derived as
\begin{equation}\label{final_result2}
	\tilde p = \sum\limits_{{k_{{n^{*}},{n^{'}}}} = 0}^G {\mathbb{P}\left\{ {{p_{{n^{*}}}} > {p_{{n^{'}}}}| \mathbf{C}} \right\}\mathbb{P}\left\{ {{K_{{n^{*}},{n^{'}}}} = {k_{{n^{*}},{n^{'}}}}} \right\}}.
\end{equation}

Following the above analysis, the heuristic performance metrics  $\tilde{p}$ in \eqref{final_result1} and \eqref{final_result2} are  functions of $L$, and the optimal codebook parameters can be conveniently calculated using closed-form expressions, i.e., $L^* = \arg \mathop {\max }\limits_L \tilde p\left( L \right)$.

\section{Simulation Results}\label{sim}
In this section, we provide numerical results to evaluate the proposed beam training method and compared with other existing schemes. We assume that the base station is equipped with $N=128$ antennas, and there are 128 DFT beams in total. The number of training beams $M=64$.

\begin{figure}[t]
	\centering
	\includegraphics[width=2.62in]{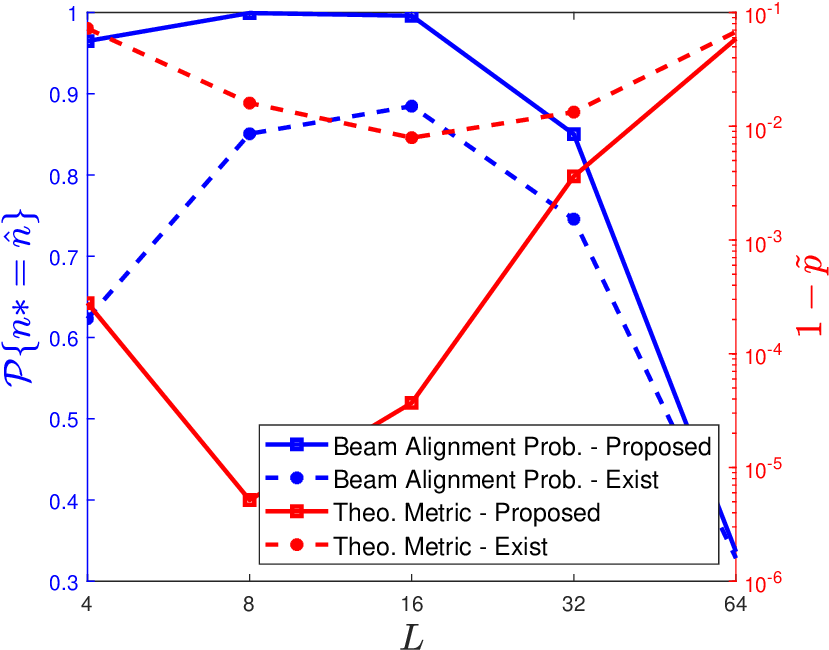}
	\caption{Relation of $L$ to beam alignment success probability and heuristic parameter.}
	\label{fig2}
\end{figure}

Fig. \ref{fig2} validates the theoretical analysis and shows the theoretical and simulated beam alignment performance of the proposed hash codebook and the random codebook for different $L$. We set the channel contains only LoS paths and lies in the grid point as described in Section \ref{sec4}, and the SNR was 10dB. In Fig. \ref{fig2}, the blue curve corresponds to the scale on the left side of the axis, indicating the successful probability of beam alignment $P\{n^{\ast}=\hat{n}\}$, the red curve corresponds to the logarithmic scale on the right side of the axis, and for clarity, we use the heuristic parameter as $1-\tilde{p}$. As shown in Fig. \ref{fig2}, the number of beams $L$ included in the training beam has a significant impact on the beam alignment performance. And the proposed heuristic parameter $\tilde{p}$ are approximately monotonic with the beam alignment performance. Therefore, we can obtain optimal $L^*$ by maximizing $\tilde{p}$.

\begin{figure}[t]
	\centering
	\includegraphics[width=2.4in]{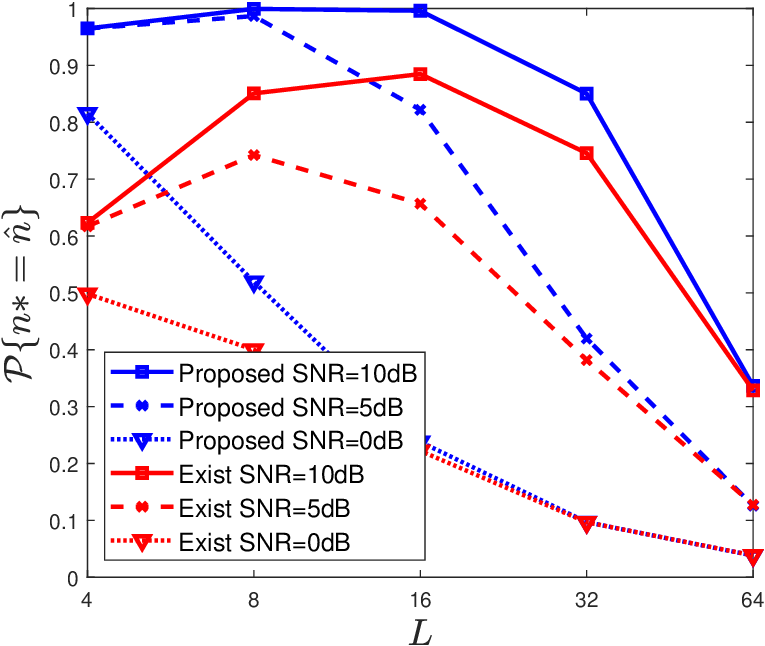}
	\caption{Relation of $L$ to beam alignment success probability in different SNR.}
	\label{fig3}
\end{figure}

Fig. \ref{fig3} demonstrates the effect of $L$ on the performance of the random codebook for different SNR. It can be seen that the optimal $L$ varies with SNR. When SNR is high, a moderate value of $L$ leads to the best performance of beam alignment. However, when the SNR is low, the optimal value of $L$ is small. This is because $L$ and SNR jointly determine the mean and variance of the Gaussian random variable $p_{n^\ast}-p_{n'}$, as shown in \eqref{approx problem}. In addition, when the parameter $L$ is properly set, the beam alignment performance of the proposed random codebook is much better than that of the completely random codebook.

\begin{figure}[t]
	\centering
	\includegraphics[width=2.4in]{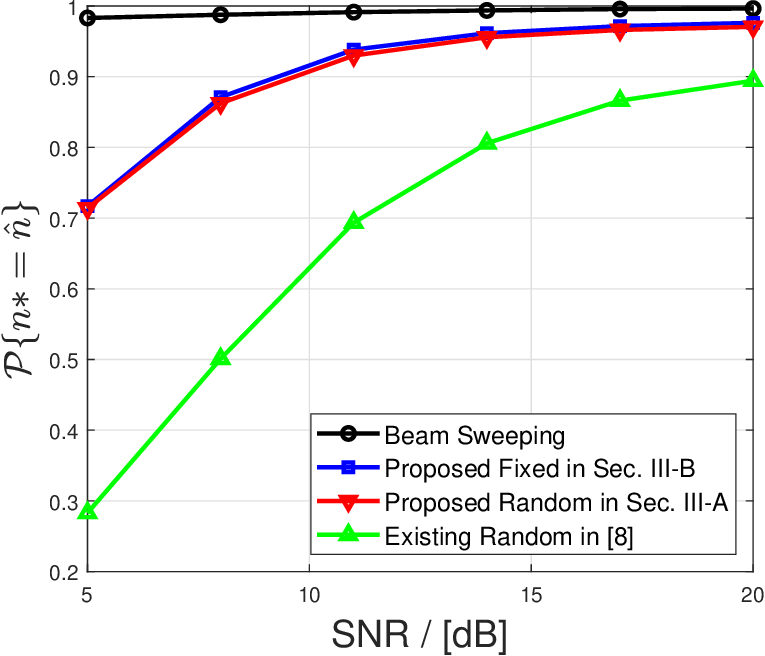}
	\caption{Beam training success rate with $N=128, M=64, L=8, P=3$.}
	\label{fig4}
\end{figure}

Fig. \ref{fig4} shows beam alignment performance of different schemes in practical terahertz channels. There are $P=3$ paths in total, containing one LoS path and two NLoS paths, where the LoS path gain $\alpha_1 \sim \mathcal{CN}(0,1)$, NLoS path gain $\alpha_2,\alpha_3\sim \mathcal{CN}(0,0.01)$. The angle of the $p$-th path $\theta_p$ follows a uniform distribution in $[-\pi/2,\pi/2]$, and the fixed codebook used for simulation is selected from 1000 random codebooks. It can be seen that the beam alignment performance of the proposed hash codebook is much better than that of the completely random codebook in references \cite{ref11}. At high SNR, the proposed hash codebook approaches the performance of beam sweeping, but requires only half the training overhead. In addition, compared with the proposed random hash codebook, the proposed fixed hash codebook can further improve the success rate of beam alignment by about 1\%, and has the advantage of easy implementation.

\section{Conclusion}\label{sec6}
In this paper, to solve the problem of high beam training overhead for extremely large antenna arrays, we study the fast beam training scheme based on hash function and analyze the impact of hash codebook on the performance of beam alignment. We propose two codebook design methods, random and fixed, and obtain the optimal codebook generation parameters according to the theoretical analysis results. Simulation results show that the proposed scheme achieves close to the performance of beam sweeping, but only requires half of the training overhead. At the same time, the accuracy of beam alignment of the proposed scheme is much better than that of the existing schemes using completely random hash codebook.

\bibliographystyle{IEEEtran}
\bibliography{mybib}
\end{document}